\documentclass[a4paper,11pt]{article}
\usepackage{pos}
\usepackage{physics}
\usepackage{siunitx}

\usepackage{lineno}
\usepackage{hyperref}
\usepackage[nooneline]{subfigure}
\subfiguretopcaptrue

\title{Development of a blue-mirror multilayer coating on light concentrators for future SiPM cameras}
 \ShortTitle{Blue-mirror Coating for SiPM Cameras}

\author*[a,b,c]{Akira~Okumura}
\author[a]{Junya~Haga}
\author[d]{Chiaki~Inoue}
\author[d]{Keiji~Nishimoto}
\author[a]{Kazuhiro~Furuta}
\author[a,b,c]{Hiroyasu~Tajima}

\affiliation[a]{Institute for Space--Earth Environmental Research, Nagoya University,\\Furo-cho, Chikusa-ku, Nagoya 464-8601, Japan}
\affiliation[b]{Kobayashi--Maskawa Institute for the Origin of Particles and the Universe, Nagoya University,\\Furo-cho, Chikusa-ku, Nagoya 464-8602, Japan}
\affiliation[c]{Nagoya University Southern Observatories, Nagoya University,\\Furo-cho, Chikusa-ku, Nagoya 464-8602, Japan}
\affiliation[d]{R\&D Department, Tokai Optical Holdings Co., Ltd., Shinpukuji-cho, Okazaki 444-2106, Japan}

\emailAdd{oxon@mac.com}

\abstract{Silicon photomultipliers (SiPMs) have a few advantages over conventional photomultiplier tubes (PMTs) used in imaging atmospheric Cherenkov telescopes. The first notable characteristic is their higher photon detection efficiency (PDE) of up to about 60\%, which is roughly 1.2--1.5 times better than that of PMTs in the 300--450\,nm range, enabling us to lower the energy threshold of gamma-ray observations and increase the photon statistics. The second advantage is that SiPMs are chemically stable after exposure to long and bright illumination, while PMTs can cause gain and quantum efficiency degradation after the same exposure. Therefore, the use of SiPMs under bright or full moon conditions may extend the total observation time in the highest energy coverage region of individual telescopes. However, the SiPM PDE is too high in wavelengths longer than 500 nm; hence, the signal-to-noise ratio (S/N) of the Cherenkov signal over the night-sky background (NSB) is not necessarily superb. This is because the Cherenkov signal is dominant over the wavelength of 300--500 nm, while the NSB is brighter in the region of 550 nm or longer. To improve the S/N with minimal and cost-effective additional hardware, we have developed multilayer coating designs with only 8 layers and applied them to the specular surfaces of light concentrators. The layers were designed to reflect more photons in the 300--500 nm range but fewer in 550--800 nm. Using a prototype light concentrator fabricated with the novel multilayer design, we demonstrated that a SiPM array exhibits ${\sim}50$\% better photon collection efficiency at 403\, nm than that obtained with PMTs, agreeing with the result of a ray-tracing simulation. The efficiency measured at 830\,nm was also successfully reduced by 30--50\%.}

\ConferenceLogo{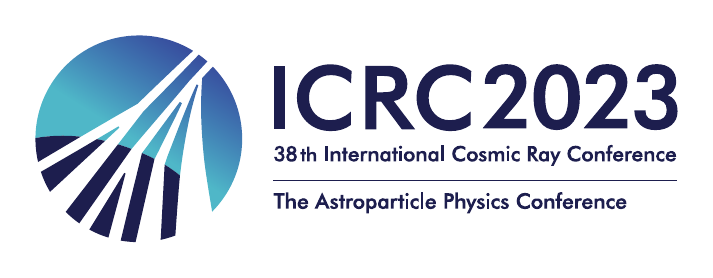}

\FullConference{%
The 38th International Cosmic Ray Conference (ICRC2023)\\
  26 July -- 3 August, 2023\\
  Nagoya, Japan}

\begin{document}
\maketitle

\section{Introduction}

\begin{figure}[b]
  \centering
  \includegraphics[width=.7\textwidth,clip]{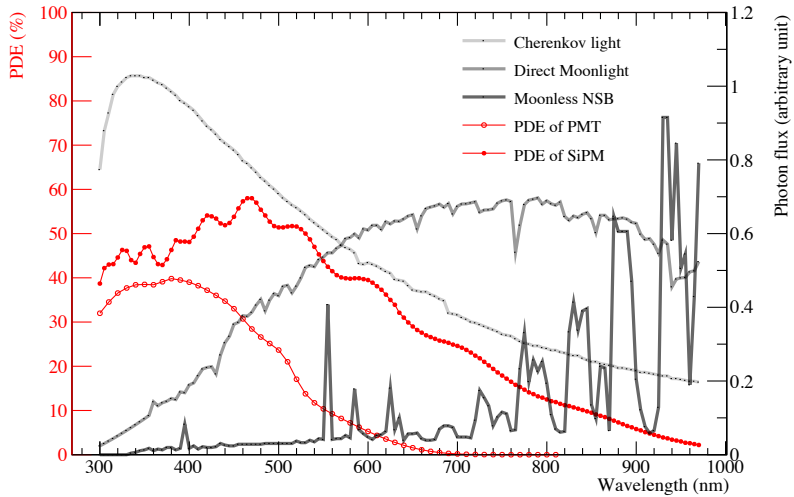}
  \caption{Typical PDE curves of LST PMTs and UV-sensitive SiPMs (the left vertical axis and red curves) and comparison of the Cherenkov light spectrum at the ground level, direct moonlight, and NSB of moonless nights (the right vertical axis and gray curves).}
  \label{fig:spectra}
\end{figure}

The use of silicon photomultipliers (SiPMs) in very-high-energy gamma-ray observations has been studied and realized in several imaging atmospheric Cherenkov telescopes (IACTs) to reduce the energy threshold and to improve the gamma-ray source detection sensitivity by utilizing a few advantages of SiPMs over conventional photomultiplier tubes (PMTs) \cite{Anderhub:2013:Design-and-operation-of-FACT--the-first-G-APD-Cher,Adams:2021:Detection-of-the-Crab-Nebula-with-the-9.7-m-protot,White:2022:The-Small-Sized-Telescopes-for-the-Southern-Site-o,Heller:2017:An-innovative-silicon-photomultiplier-digitizing-c,Lombardi:2020:First-detection-of-the-Crab-Nebula-at-TeV-energies}. First, as illustrated in Fig.~\ref{fig:spectra}, the photon detection efficiencies (PDEs) of SiPMs sensitive to ultraviolet (UV) photons are comparable or higher than those of UV-sensitive PMTs in the 300--500\,nm range with an avalanche photodiode (APD) cell size of 50\,$\mu$m or 75\,$\mu$m. Second, SiPMs have compact pixel sizes, typically from $1\times1$\,mm$^{2}$ to $6\times6$\,mm$^{2}$, which are suitable for building compact and wide field-of-view (FOV) cameras comprising a few thousand pixels or larger cameras with finer pixel resolution. Third, SiPMs are more tolerant to bright moon conditions, allowing longer observation times for monitoring and deep surveys even though the trigger threshold becomes higher owing to increased night sky background (NSB).

Despite the aforementioned advantages of SiPMs, a few limitations make them unsuitable for very-high-energy gamma-ray observations. For instance, the PDE of UV-sensitive SiPMs is unnecessarily too high in the wavelength range of 550\,nm and longer, where the Cherenkov photon spectrum (photon number density per wavelength) decreases ($\mathrm{d}{N}/\mathrm{d}\lambda \propto 1/\lambda^2$) but the amount of NSB rapidly increases because of airglow lines, as compared in Fig.~\ref{fig:spectra}. This implies that the first advantage of SiPMs, namely, having higher PDE than that of PMTs, can be easily canceled out by the NSB in the absence of any measure to block or reduce the long-wavelength photons, thereby resulting in no improvement in the trigger or analysis thresholds of Cherenkov images.

To circumvent a possible NSB contamination, several approaches have previously been proposed to cut photons with wavelengths longer than ${\sim}500$\,nm. For example, a large-aperture UV-pass filter was manually installed in front of the PMT array of the MAGIC-1 camera to absorb the NSB and scattered photons from the Moon. This enabled MAGIC~1 to perform gamma-ray observations even under bright moon conditions \cite{Ahnen:2017:Performance-of-the-MAGIC-telescopes-under-moonligh}. However, fabricating a durable glass UV-pass filter with a diameter of ${\sim}2$\,m is expensive, and an automated filter mounting system, which is required for different sky conditions, is difficult to implement on the focal plane camera. Hence, using large UV-pass filters is not suitable for daily observations of larger IACT arrays.

Another approach for NSB rejection in IACTs is to apply a multilayer coating on the mirror surface or camera window. As regards the Small-Sized Telescopes (SSTs) of the Cherenkov Telescope Array (CTA), the next-generation ground-based gamma-ray observatory \cite{Acharya:2013:Introducing-the-CTA-concept,Actis:2011:Design-concepts-for-the-Cherenkov-Telesc}, a few-layer coating on the primary and secondary mirrors has been studied \cite{Palombara:2022:Mirror-production-for-the-Cherenkov-telescopes-of-} to enhance the UV--blue reflectance and slightly decrease the reflectance at longer wavelengths. A multilayer coating on a Borofloat camera window installed in front of the photodetectors has also been studied to selectively reflect back photons with wavelengths longer than 550\,nm \cite{Alispach:2020:Large-scale-characterization-and-calibration-strat}. 

The CTA SSTs will employ SiPM arrays for the focal planes of the cameras \cite{White:2022:The-Small-Sized-Telescopes-for-the-Southern-Site-o}. This is because the main physics goal of the SSTs is to study galactic PeVatrons in the energy range of ${\sim}5$--300\,TeV by spreading 70 telescopes with a 4\,m primary mirror and ${\sim}9^\circ$ FOV. A compact focal plane with a diameter of ${\sim}40$\,cm, covered with 2048 of $6\times6$\,mm$^2$ SiPMs, is suitable for such an optical system. A multilayer coating with a few and several tens of layers for the relatively small mirror and window surfaces, respectively, is a cost-effective solution.

In contrast, the CTA Large- and Medium-Sized Telescopes (LSTs and MSTs) will have an array of ${\sim}1800$ PMTs on the focal plane, where hexagonal light concentrators (often referred to as Winston cones) are attached to individual PMTs to minimize the dead area of the circular PMTs and reduce stray NSB coming from directions other than those containing mirrors. Owing to the camera size being larger than 2\,m and the acrylic window material, the absorptive UV-pass filter approach or applying multilayer coating on the camera window is infeasible. Therefore, other technical solutions must be investigated beforehand if a future camera upgrade of LSTs or MSTs is planned by replacing PMTs with SiPMs.

In this paper, we propose novel multilayer coating designs that absorb more than 50\% of photons with wavelengths longer than ${\sim}550$\,nm. By applying the coating on the specular surfaces of light concentrators, rather than on the camera window, it can reduce the NSB hitting the focal plane with a diameter of ${\sim}2$\,m. The coating solely comprises 8 layers to avoid high production costs while maintaining the reflectance in 300--550\,nm and cutoff at around 550\,nm as sharp as possible.

\section{SiPM Cameras for Large-Sized Telescopes}

The CTA will have three different IACT designs to have a vast effective area of up to a few square kilometers and wide energy coverage from 20\,GeV to 300\,TeV. This will be achieved by building three different telescope designs, of which the lowest energy part, 20\,GeV to 3\,TeV will be observed by the LSTs. The LSTs will achieve this low energy threshold by using segmented parabolic systems with an effective diameter of 23\,m and focal plane camera comprising 1885 hexagonal pixels with an angular size of $0.1^\circ$ (side to side).

Owing to the parabolic optical system, which yields the best point spread function (PSF) at the camera center, the first LST built at the Observatorio del Roque de los Muchachos, Spain, has achieved an on-axis PSF size that is smaller than the aforementioned pixel size by a factor of ${\sim}1.5$. Therefore, to fully exploit the optical performance of the LST optics, the trigger and analysis threshold can be further reduced and Cherenkov image analysis can be improved by replacing the LST cameras with smaller-pixel cameras in the future.

A working package in the CTA-LST project is currently conducting a feasibility study on a SiPM camera to be implemented as a future upgrade of the LST cameras \cite{Matthieu:2023:The-next-generation-cameras-for-the-Large-Sized-Te}. Investigating the novel multilayer coating design for light concentrators in this study is and activity in the working package. Some of the other technical and simulation studies are also presented in these proceedings \cite{Matthieu:2023:The-next-generation-cameras-for-the-Large-Sized-Te,Saito:2023:Characterization-of-SiPM-and-development-of-test-b}.

\section{Multilayer Design}

To design a multilayer coating for the specular surfaces of LST light concentrators, the following requirements must be considered. First, the reflectance curve must have a cutoff at around 550\,nm to reduce the NSB contamination; however, the reflectance in the 300--550\,nm range must be comparable to or higher than that of a simple UV-enhanced aluminum coating. Second, the reflectance performance should be optimized for angles of incidence around 65$^\circ$ and should not be highly dependent on the angles. This is because photons coming from the mirror directions hit the light concentrator surfaces at angles from ${\sim}40^\circ$ to $90^\circ$ \cite{Okumura:2017:Prototyping-hexagonal-light-concentrators-using-hi}. Third, the number of layers should not exceed ${\sim}10$ and only readily available materials, such as SiO$_2$ and Ta$_2$O$_5$, should be used. This is to reduce the coating cost and easily control the layer thickness.

To optimally satisfy these requirements, we have designed 8-layer coatings, where a thin aluminum layer is inserted to make the coating absorptive \cite{Okumura::Japanese-Patent-Application-No.-2021-31974-filed-o}. Table~\ref{table:design} compares the multilayer design used for the first LST and our 3 novel designs with different aluminum thicknesses (5, 10, and 15\,nm) in layer 5.

% Table taken from Nishimoto-san's e-mail on 2020/7/6 
\begin{table}
  \centering
  \caption{Comparison of multilayer designs. Designs 1--3, which are proposed in this study, have 8 layers on the ABS substrate and an adhesion layer made of 10\,nm AlO$_2$. The designs are identical except for the aluminum thickness in layer 5. In contrast, the LST-1 design has only three layers to form a simple high-reflector coating with low- and high-index materials on the aluminum layer.}
\begin{tabular}{llllrrrr}
  \hline
  & Material & $(n, k)$ @ 400\,nm & Design 1 & Design 2 & Design 3 & LST-1 Design \\
  & & & (nm) & (nm) & (nm) & (nm) \\
  \hline
  \hline
             & Air         &          &          &          &              \\
  \hline
Layer 8 & SiO$_2$     & (1.47, 0.00) & 35.93    & 35.93    & 35.93    & ---          \\
Layer 7 & Ta$_2$O$_5$ & (2.20, 0.00) & 35.72    & 35.72    & 35.72    & ---          \\
Layer 6 & SiO$_2$     & (1.47, 0.00) & 39.14    & 39.14    & 39.14    & ---          \\
Layer 5 & Al          & (0.38, 4.23) & 5.00     & 10.00    & 15.00    & ---          \\
Layer 4 & SiO$_2$     & (1.47, 0.00) & 41.88    & 41.88    & 41.88    & ---          \\
Layer 3 & Ta$_2$O$_5$ & (2.20, 0.00) & 42.49    & 42.49    & 42.49    & 20.11        \\
Layer 2 & SiO$_2$     & (1.47, 0.00) & 68.97    & 68.97    & 68.97    & 72.56        \\
Layer 1 & Al          & (0.38, 4.23) & 123.00   & 123.00   & 123.00   & 112.59       \\
  \hline
Layer 0 & AlO$_2$   &  & 10.0     & 10.0     & 10.0     & 10.0        \\
Substrate    & ABS       &  &          &          &          &             \\
  \hline
\end{tabular}
\label{table:design}
\end{table}

\begin{figure}
  \centering
  \makebox[\textwidth][c]{
  \subfigure[]{%
    \includegraphics[width=.6\textwidth,clip]{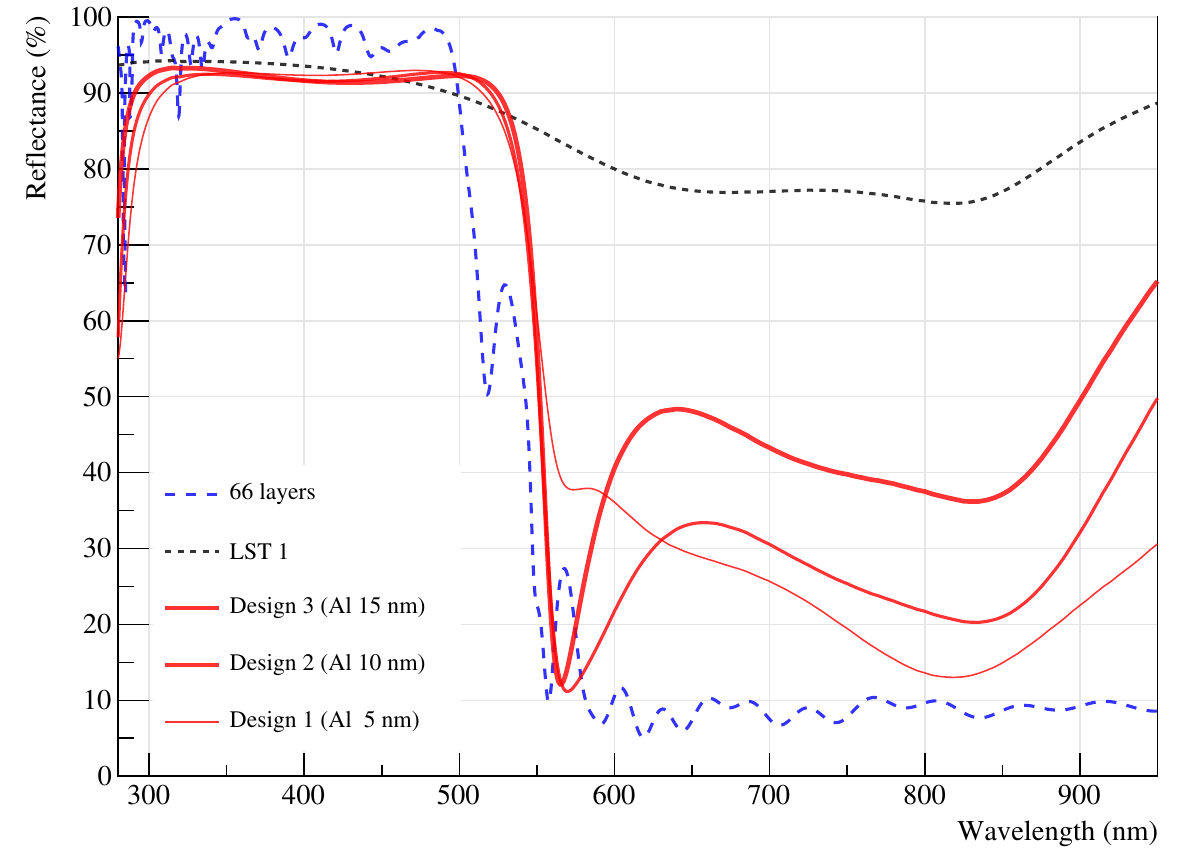}
    \label{fig:design65}
  }%
  \subfigure[]{%
    \includegraphics[width=.6\textwidth,clip]{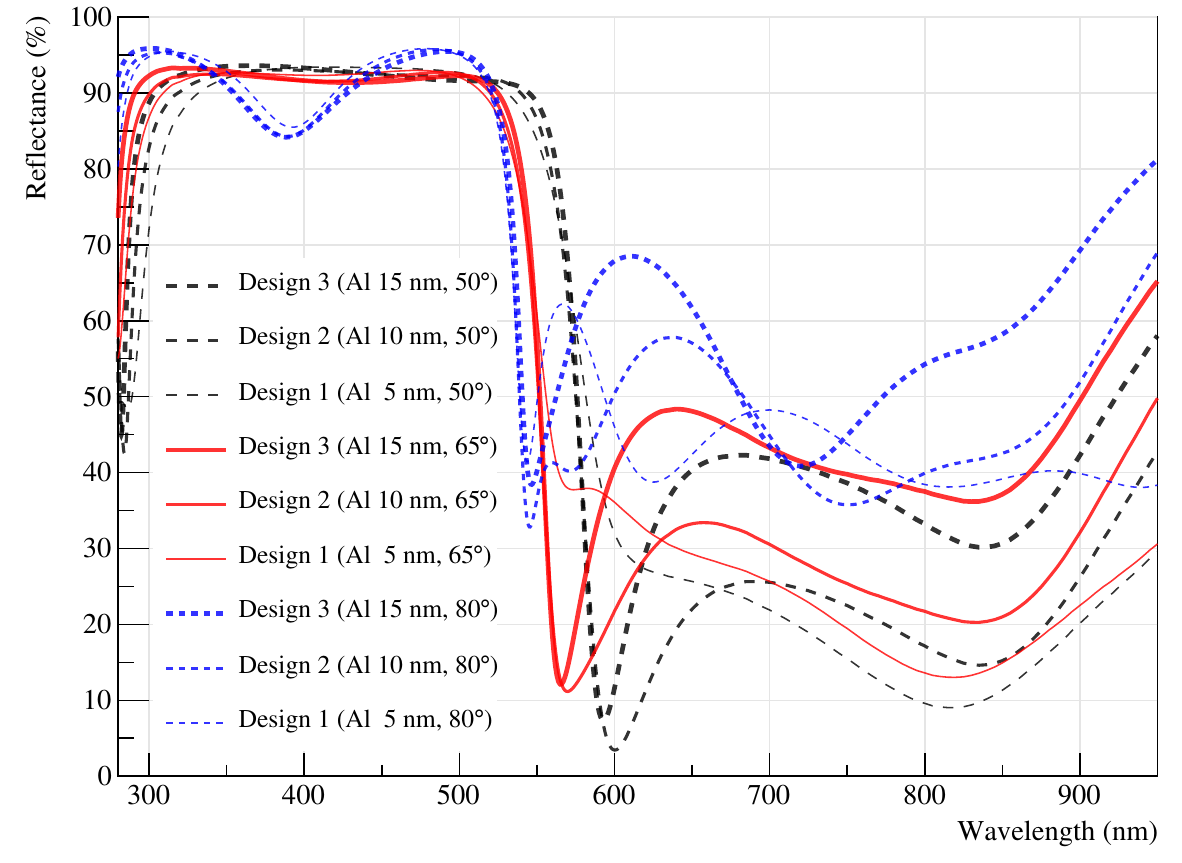}
    \label{fig:design50_80}
  }
  }
  \caption{(a) Simulated reflectance curves as a function of wavelength for an angle of incidence fixed at 65$^\circ$. Five different multilayer designs are compared. (b) Simulated reflectance curves of Designs 1--3 with three different angles, 50$^\circ$, 65$^\circ$, and 80$^\circ$. These simulations were performed using the ROBAST library \cite{Okumura:2022:ROBAST3,Okumura:2016:ROBAST:-Development-of-a-ROOT-based-ray-tracing-li}.}
\end{figure}

Fig.~\ref{fig:design65} compares the simulated reflectance values of Designs 1--3 presented in Table~\ref{table:design}, assuming an angle of incidence fixed at $65^\circ$. The values for the LST-1 coating and a 66-layer coating are also presented as references. Reflectance curves for three different angles are compared in Fig.~\ref{fig:design50_80}. Evidently our novel designs can reduce the NSB by approximately half for wavelengths longer than ${\sim}550$\,nm for different angles, while the 66-layer coating exhibits higher reflectance in the 300--500\,nm range and lower reflectance in the 600--900\,nm range.

\section{Prototyping and Measurements}

\begin{figure}
  \centering
  \makebox[\textwidth][c]{
  \subfigure[]{%
  \includegraphics[width=.43\textwidth,clip]{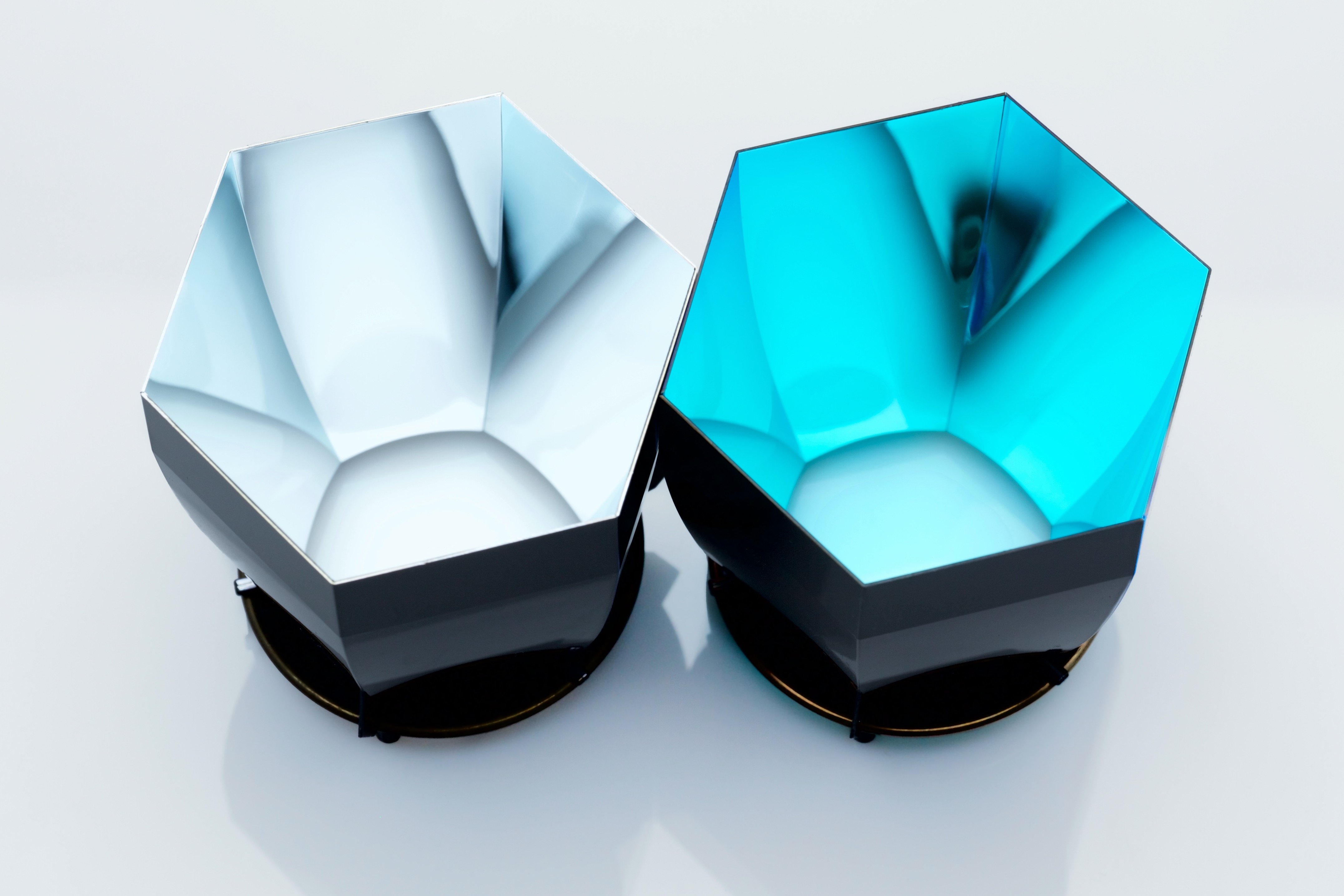}
    \label{fig:cones}
  }%
  \subfigure[]{%
    \includegraphics[width=.285\textwidth,clip]{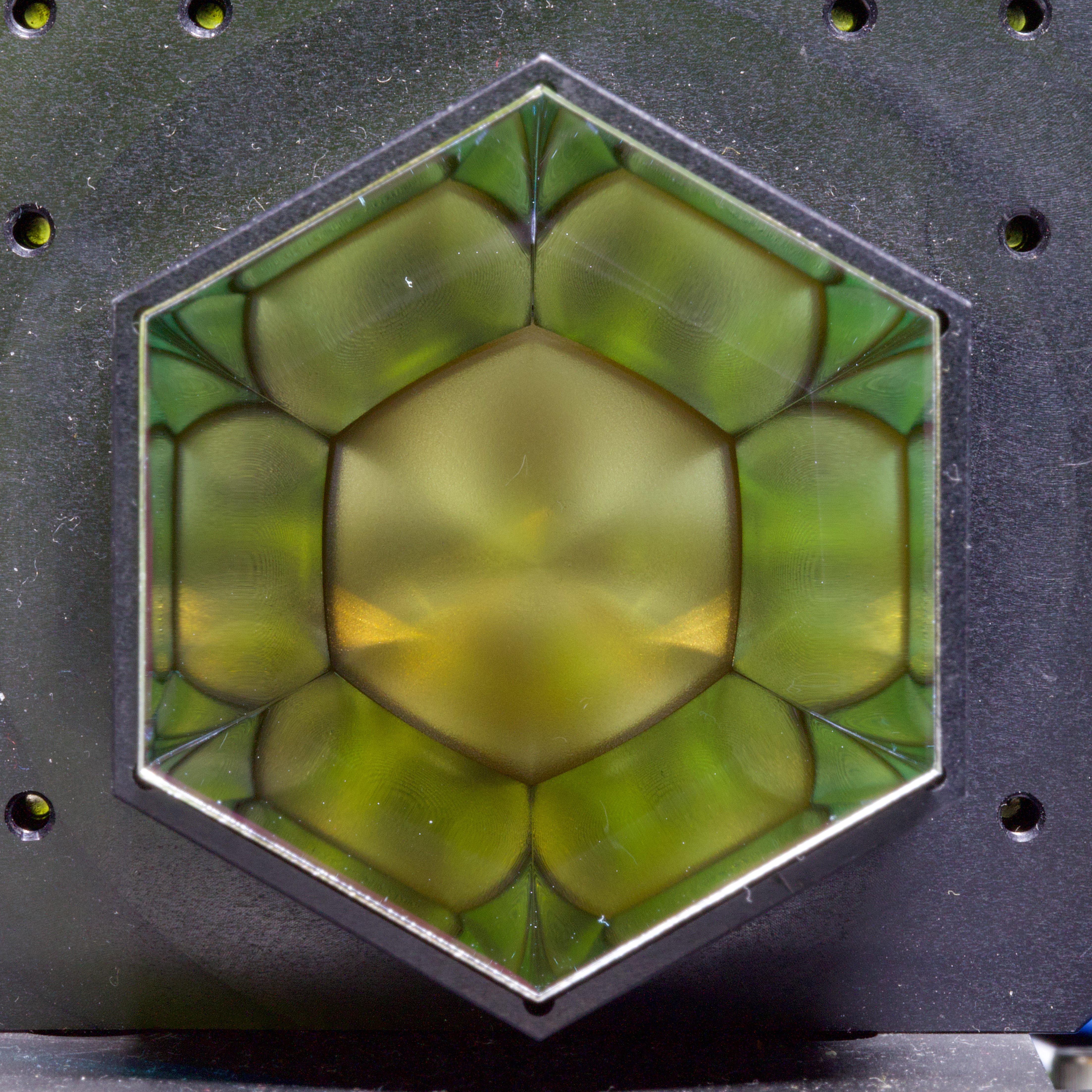}
    \label{fig:cone_with_PMT}
  }%
  \subfigure[]{%
    \includegraphics[width=.285\textwidth,clip]{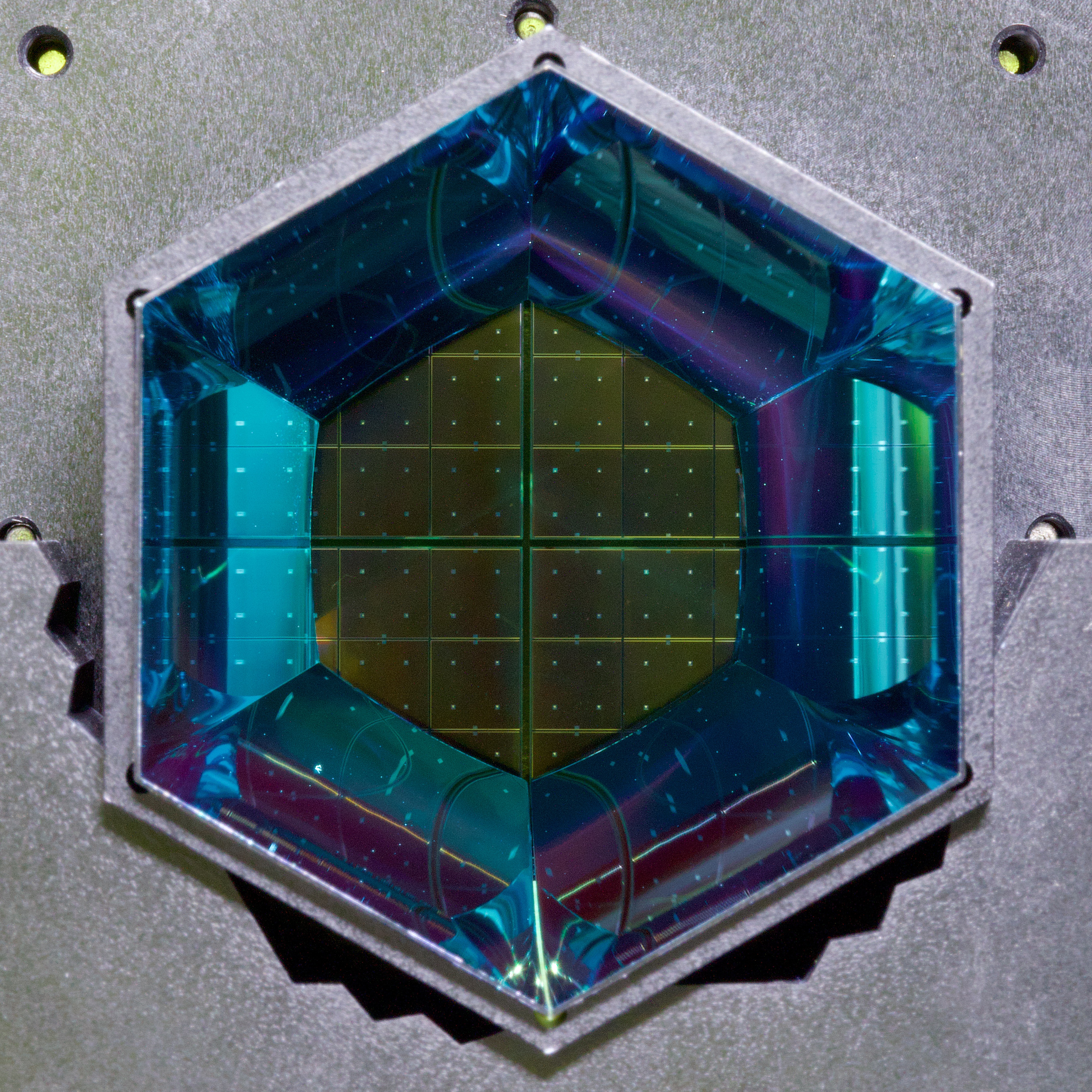}
    \label{fig:cone_with_SiPM}
  }
  }
  \caption{(a) Photograph of a hexagonal light concentrator for LSTs (left, LST-1 cone) and one with a coating created according to Design 3 (right, blue cone). (b) LST-1 cone coupled to a PMT. (c) Blue cone coupled to an SiPM array.}
\end{figure}

We have prototyped a light concentrator for SiPMs by applying the multilayer Design 3 to an uncoated LST-1 cone. Fig.~\ref{fig:cones} compares a normal LST-1 cone and our prototype. Each cone was coupled with an LST-1 PMT (Hamamatsu Photonics R11920- 100-20) or a SiPM array (Hamamatsu Photonics S14521-8649, APD cell size 75\,$\mu$m) (Fig.~\ref{fig:cone_with_PMT} and \ref{fig:cone_with_SiPM}), and their collection efficiencies were measured using the method reported in \cite{Okumura:2017:Prototyping-hexagonal-light-concentrators-using-hi} and plotted as a function of angle of incidence at the focal plane.

\begin{figure}
  \centering
  \includegraphics[width=.6\textwidth,clip]{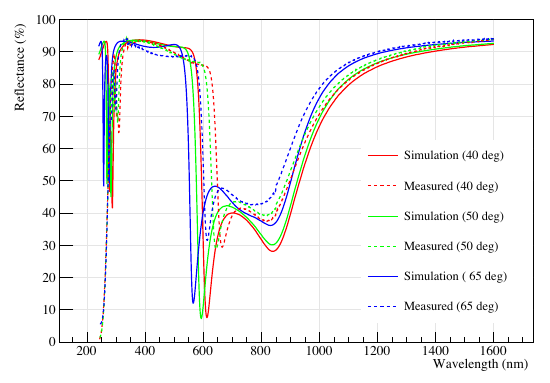}
  \caption{Simulated reflectance curves of Design 3 compared to  measurements.}
  \label{fig:measured_spectra}
\end{figure}

Design 3 does not yield the lowest reflectance at long wavelengths, as shown in Figure~\ref{fig:design50_80}. However, its actual performance is better than that of the other two designs because the actual optical behavior of a 5--10\,nm Al layer differs from that depicted in an ideal simulation. Fig.~\ref{fig:measured_spectra} compares simulated and measured reflectance curves for three different angles. The measured curves are shifted to longer wavelengths by ${\sim}50$\,nm; the reason needs to be investigated.

\begin{figure}
  \centering
  \makebox[\textwidth][c]{
  \subfigure[]{%
    \includegraphics[width=.6\textwidth,clip]{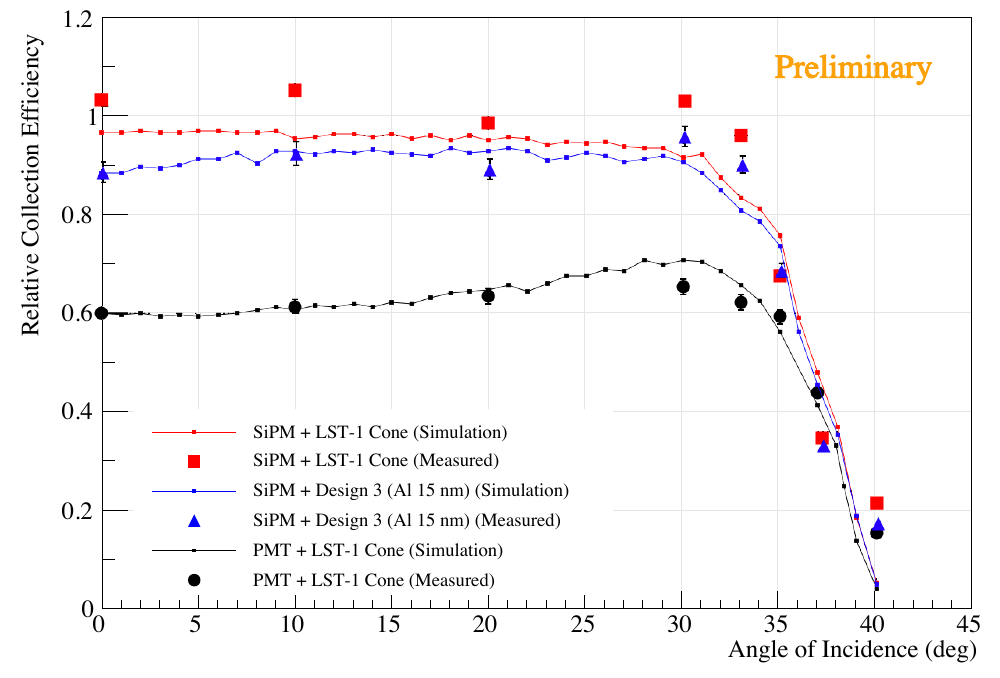}
    \label{fig:403nm}
  }%
  \subfigure[]{%
    \includegraphics[width=.6\textwidth,clip]{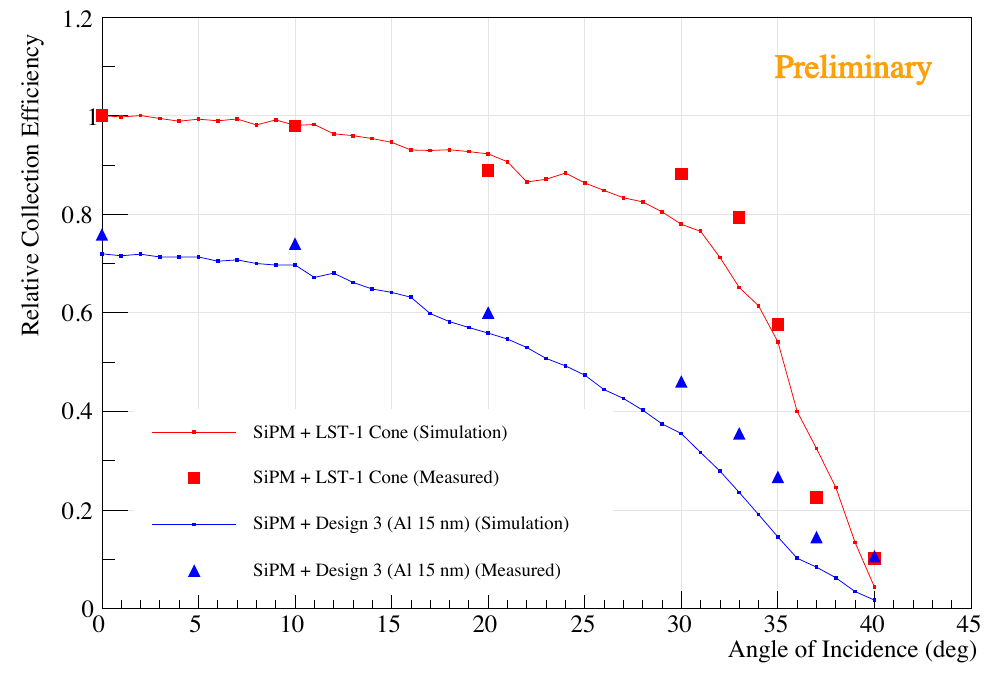}
    \label{fig:830nm}
  }
  }
  \caption{(a) Comparison of relative collection efficiencies of three different configurations at 403\,nm: An SiPM array and LST-1 cone (red), same but the coating is implemented according to Design 3 (blue), and an LST-1 PMT and LST-1 cone (black). Measurements (symbols) and ROBAST simulations (lines) are compared. (b) Same as (a) but the input wavelength is 830\,nm at which the PMT is not sensitive.}
\end{figure}

Fig.~\ref{fig:403nm} and \ref{fig:830nm} demonstrate the relative collection efficiencies of three different configurations, namely, SiPM and LST-1 cone, SiPM and Design 3, and PMT and LST-1 cone. ROBAST simulations are compared with measurements taken at two different wavelengths: 403\,nm and 830\,nm.

Using a SiPM array instead of an LST PMT increased the relative collection efficiency by ${\sim}50$\% at 403\,nm, thus demonstrating that SiPMs are more sensitive to the Cherenkov signal compared to PMTs. However, the collection efficiency achieved by Design 3 was ${\sim}10$\% worse than that of the normal LST-1 cone. In contrast, at 830\,nm, the collection efficiency of the Design-3 cone was 30--50\% lower than that of the normal LST-1 cone. Therefore, by further improving reflectance in the 300--500\,nm range and tuning the cutoff, our novel multilayer coating and SiPMs will provide a better signal-to-noise ratio than the existing LST-1 cone coating.

\section{Conclusion}

We have developed novel multilayer coating designs for future SiPM cameras used in IACTs. The first prototype light concentrators were fabricated using one of our designs, yielding a collection efficiency better than that of the current LST-1 camera pixels at 403\,nm. At 830\,nm, the efficiency was lower as expected from a simulation; this is crucial for suppressing the NSB contamination of the Cherenkov signals. Further coating optimization and measurements at other wavelengths are planned and will be reported later.

\providecommand{\href}[2]{#2}\begingroup\raggedright\endgroup

\acknowledgments

This study was supported by JSPS KAKENHI Grant Numbers JP18KK0384, JP20H01916, and JP23H04897. We also acknowledge the financial support for the CTA-LST project listed on \url{https://www.lst1.iac.es/acknowledgements.html}.

%% Full authors list (ONLY FOR COLLABORATIONS)
%\clearpage
%\section*{Full Authors List: \Coll\ Collaboration}
%
%\noindent \textbf{Note comment afterwards:} Collaborations have the possibility to provide an authors list in xml format which will be used while generating the DOI entries making the full authors list searchable in databases like Inspire HEP. For instructions please go to icrc2021.desy.de/proceedings or contact us under icrc2021proc@desy.de.\\
%
%\scriptsize
%\noindent
%first.author$^1$, 
%second.author$^2$, 
%third.author$^3$ % .... more names
%and 
%last.author$^{n}$ \\
%
%\noindent
%$^1$first.affiliation.
%$^2$second.affiliation. % .... more affiliation
%$^{m}$last.affiliation.

\end{document}